\documentclass[aps,prb,floatfix,showpacs,twocolumn]{revtex4}
\usepackage[english]{babel}
\usepackage{graphicx}
\usepackage{dcolumn}
\usepackage{amsmath}

\newcommand{\vs}{{\it vs.}}
\newcommand{\eg}{e.g.}
\newcommand{\ie}{i.e.}
\newcommand{\etal}{{\it et al.}}
\newcommand{\alphaETI}{$\alpha$-(BEDT-TTF)$_2$I$_3$}
\newcommand{\cm}{cm$^{-1}$}
\newcommand{\ud}{\mathrm{d}}

\begin{document}

\title{Electrodynamic Response of the Charge Ordering Phase: \\
Dielectric and Optical Studies of $\alpha$-(BEDT-TTF)$_2$I$_3$}

\author{T.\ Ivek}
\email{tivek@ifs.hr}
\homepage{http://real-science.ifs.hr/}
\author{B.\ Korin-Hamzi\'{c}}
\author{O.\ Milat}
\author{S.\ Tomi\'{c}}
\affiliation{Institut za fiziku, P.O.Box 304, 10001 Zagreb, Croatia}
\author{C.\ Clauss}
\author{N.\ Drichko}
\author{D.\ Schweitzer}
\author{M.\ Dressel}
\affiliation{Physikalisches Institut, Universit\"at Stuttgart, Pfaffenwaldring 57, 70550 Stuttgart, Germany}

\date{\today}

\begin{abstract}
We report on the anisotropic response, the charge and lattice dynamics of normal 
and charge-ordered phases with horizontal stripes in single crystals of the 
organic conductor \alphaETI{} determined by dc resistivity, dielectric 
and optical spectroscopy. An overdamped Drude response and a small conductivity 
anisotropy observed in optics is consistent with a weakly temperature 
dependent dc conductivity and anisotropy at high temperatures. The splitting of 
the molecular vibrations $\nu_{27}(\mathrm{B}_u)$ evidences the abrupt onset of 
static charge order below $T_\mathrm{CO}=136$~K. The drop of optical 
conductivity measured within the $ab$ plane of the crystal is characterized by 
an isotropic gap that opens of approximately 75~meV with several phonons 
becoming pronounced below. Conversely, the dc conductivity anisotropy rises
steeply, attaining at 50~K a value 25 times larger than at high temperatures.
The dielectric response within this plane reveals two broad relaxation modes of 
strength $\Delta\varepsilon_\mathrm{LD} \approx 5000$ and 
$\Delta\varepsilon_\mathrm{SD} \approx 400$, centered at 
$1~{\rm kHz}<\nu_\mathrm{LD}<100$~MHz and $\nu_\mathrm{SD} \approx 1$~MHz. The 
anisotropy of the large-mode (LD) mean relaxation time closely follows the 
temperature behavior of the respective dc conductivity ratio. We argue that this 
phason-like excitation is best described as a long-wavelength excitation of a 
$2k_\mathrm{F}$ bond-charge density wave expected theoretically for layered 
quarter-filled electronic systems with horizontal stripes. Conversely, based on 
the theoretically expected ferroelectric-like nature of the charge-ordered phase, 
we associate the small-mode (SD) relaxation with the motion of domain-wall 
pairs, created at the interface between two types of domains, along the $a$ and 
$b$ axes. We also consider other possible theoretical interpretations and 
discuss their limitations.
\end{abstract}


%
%
%
%
%
%
%

\maketitle

\section{Introduction}
The competition between the tendency of electrons to delocalize and the presence 
of electron-electron interaction, electron-phonon interaction -- often combined 
with spin-phonon interaction -- is the origin of the extremely rich phase 
diagrams in the condensed matter systems with reduced 
dimension.\cite{Fulde93,MC04} Among the most intriguing phenomena found in 
these systems are broken-symmetry phases like charge- and spin-density waves 
(CDW, SDW), charge orders (CO), antiferromagnetic, spin-Peierls phases and 
superconductivity. These phases show a large variety of nonlinear properties and 
complex dynamics, including collective 
excitations.\cite{Gruener88,Littlewood87,Cava85,VuleticPR,Dressel09}. While the 
conventional (2$k_\mathrm{F}$) CDW implies periodically modulated charge density 
and is caused by the electron-phonon interaction, the $4k_\mathrm{F}$ charge 
modulation of Wigner-crystal type, often called charge ordering, originates from 
strong on-site Coulomb repulsion $U$. Thus, the CO is most commonly thought of 
as the alternating localized charge of different valencies arranged in a crystal 
lattice. In particular, a wealth of CO phenomena with charge disproportionation 
is found in organic conductors characterized by different anisotropic networks. 
These include quasi-one-dimensional (TMTTF)$_2X$ (Ref.\ \onlinecite{Monceau01}), 
(DI-DCNQI)$_2$Ag (Ref.\ \onlinecite{Kanoda07}) and quasi-two-dimensional 
conductors based on the BEDT-TTF [bis(ethylenedithio)tetrathiafulvalene]
molecule: such as $\theta$-(BEDT-TTF)$_2$RbZn(SCN)$_4$ and 
\alphaETI{}.\cite{Takahashi06} It is notable that these systems have a
quarter-filled conduction band and due to that not even the large values of
on-site repulsion $U$ with respect to the electron hopping $t$ are sufficient to 
transform the ground state from metallic to insulating. Rather, the inter-site 
Coulomb interaction $V$ is required to stablize a Wigner-crystal type 
phase.\cite{Seo06} Although CO features seem to point directly toward full 
localization of charges and their alternating arrangements, some theoretical 
studies dispute this conventional view indicating that a delocalized CDW picture 
might also be relevant to some CO, \eg{}, like the one in
$\theta$-(BEDT-TTF)$_2X$ materials.\cite{Clay02}

\begin{figure} 
\includegraphics[clip,width=0.9\columnwidth]{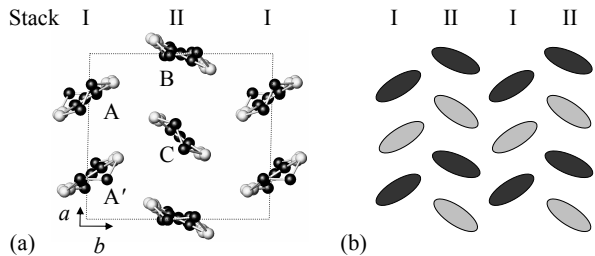}
\caption{\label{fig:a-ET-CO}(a) Schematic representation of donor layer in 
\alphaETI{}. Molecular sites belonging to the stack I and stack II are denoted 
as A, A$^\prime$ and B, C, respectively. (b) Stripe arrangement in the $ab$ 
conducting donor layer of \alphaETI{} in the charge-ordered state. Dark- and 
light-gray ovals denote charge-rich and charge-poor molecules, respectively. 
The stripes extend along the $b$-direction.}
\end{figure}

\alphaETI{}, the first organic material to show highly conductive properties in 
two dimensions,\cite{Bender84-108} is also one of the most prominent examples of 
charge order among 2D organic conductors. This system displays a rich 
temperature-pressure phase diagram with a number of intriguing quantum phenomena 
ranging from superconductivity,\cite{Tajima02}, charge order,\cite{Takahashi06} 
persistent photoconduction,\cite{Tajima05} photoinduced phase 
transition,\cite{Iwai07} nonlinear ultrafast optical response,\cite{Yamamoto08} 
and zero-gap semiconductivity\cite{Mori10} characterized by Dirac cones and
massless Dirac fermions.\cite{Tajima07,Tajima09} In addition, recently a first
successful observation was reported of correlated electron motion in the
light-matter interaction leading to the photo-induced insulator-to-metal
transition.\cite{Kawakami10}

The triclinic crystal structure is an alternation of insulating anion (I$_3^-$) 
layers and conducting layers of donor molecules (BEDT-TTF$^{0.5+}$ on average). 
The BEDT-TTF molecules form a herring-bone structure and are organized in a 
triangular lattice with two types of stacks. At room temperature stack I is 
weakly dimerized and composed of crystallographically equivalent molecules A and 
A$^\prime$, while the stack II is a uniform chain composed of B and C molecules 
[see Fig.\ \ref{fig:a-ET-CO}(a)]. Thus the unit cell contains four BEDT-TTF 
molecules. At high temperatures the system is a semimetal with small electron 
and hole pockets in the Fermi surface.\cite{Bender84,MoriCL84} A slight but 
noticable charge disproportionation is observed already at room temperature, 
indicating that CO gradually develops as the temperature is reduced towards 
$T_\mathrm{CO}$.\cite{Kakiuchi07,Moroto04}. As demonstrated by nuclear magnetic 
resonance (NMR)\cite{Takano01} and synchrotron x-ray diffraction 
measurements,\cite{Kakiuchi07} charge order at long length scales develops fully 
below the metal-to-insulator phase transition $T_\mathrm{CO} = 136$~K. At 
$T_\mathrm{CO}$ the conductivity drops by several orders of magnitude and a 
temperature-dependent gap opens in charge and spin sector which indicates 
insulating and diamagnetic nature of the ground state. x-ray diffraction 
measurements indicate subtle structural changes at $T_\mathrm{CO}$. There are no 
translations of molecules, only a shift in dihedral angles (angles between 
molecules of two neighboring stacks) is observed. This results in breaking of 
inversion symmetry between A and A$^\prime$ sites, with space group changing 
from P$\bar{1}$ to the P1, and also allows for crystal twinning in the
low-temperature acentric structure. The variations in dihedral angles cause an 
appreciable 2D modulation of overlap integrals between the BEDT-TTF 
sites.\cite{Kakiuchi07} Finally, molecular deformations are observed which are 
at the origin of charge disproportionation. Estimated charge values $\rho$ of 
the molecules A, A$^{\prime}$, B and C are $0.82(9)e$, $0.29(9)e$, $0.73(9)e$ 
and $0.26(9)e$, respectively. The exact site assignment is not settled yet, 
since these values differ slightly from those found in the NMR, vibrational 
infrared and Raman spectroscopy, and anomalous x-ray diffraction 
measurements.\cite{Takano01,Moldenhauer93,Dressel04,Drichko09,Wojciechowski03,Kakiuchi07}
Nevertheless, all these experiments consistently indicate that the 
charge order comprises ``horizontal'' charge stripes of charge-poor (CP) sites, 
the A$^{\prime}$ and C molecules, and charge-rich (CR) sites (A and B molecules) 
along the $b$ crystallographic axis, as depicted in Fig.\ \ref{fig:a-ET-CO}(b). 
Contrary to the conventional view of the CO as an alternation of localized
charges, experimental x-ray data indicate that a delocalized CDW-like picture 
might be a more appropriate description of the CO stabilized in this system.

There have been some previous attempts to find evidence of collective response 
in \alphaETI{}. More than fifteen years ago, preliminary 
measurements\cite{Dressel94} indicated the existence of a broad relaxation in 
radio-frequency range with a large dielectric constant of the order of $10^5$ as 
well as sample-dependent nonlinearities. However, no conclusive interpretation 
could be given at that time. These experiments need to be revisited and more 
detailed ones with electric fields along different crystallographic directions 
have to be added in order to shed more light on the electrodynamics of the 
charge-ordered state. Recently anisotropic voltage oscillations have been 
observed that are associated with nonlinear conductivity.\cite{Tamura10} Such 
a phenomenon closely resembles the collective sliding motion of density waves in 
quasi-one-dimensional conductors,\cite{Gruener88} which makes it yet another 
result to raise the question of similarities and differences between 
conventional density waves and CO found in 2D systems. At the time of writing of 
this paper, Yue \etal{}\cite{Yue10} presented a new detailed 
infrared and Raman study in the metallic and CO phase of \alphaETI{}.

In an attempt to clarify several issues mentioned above we have undertaken an 
investigation of electrodynamic response in the normal phase and ground state of 
the layered organic conductor \alphaETI{}. We have performed comprehensive 
optical investigations in all three crystallographic directions, as well as dc 
and ac conductivity-anisotropy measurements on carefully oriented single 
crystals of \alphaETI{}. A brief report of this investigation has previously 
been published.\cite{Ivek10} Our results demonstrate a complex and anisotropic 
dispersion in the charge-ordered state in contrast to an almost isotropic and 
temperature-independent charge response at high temperatures. The abrupt onset 
of static charge order below $T_\mathrm{CO}=136$~K is signaled by a splitting of 
molecular vibrations $\nu_{27}(\mathrm{B}_u)$ and a dramatic drop of the optical 
conductivity. Its anisotropy does not change much in contrast to anisotropy of 
dc conductivity which rises steeply with decreasing temperature. Charge 
redistribution detected in the CO phase corresponds nicely to the one estimated 
from the x-ray data.\cite{Kakiuchi07} Similar to the Peierls CDW state, we 
observe long-wavelength charge excitations with an anisotropic phason-like 
dispersion, which surface as broad screened relaxation modes along the $a$- and 
$b$-axes of the BEDT-TTF planes. In addition, we detect short-wavelength charge 
excitations in the form of domain-wall pairs, created due to inversion symmetry 
breaking, which are less mobile and induce a much weaker polarization, again 
along both crystallographic axes. Possible theoretical interpretations are 
discussed and arguments are given that the nature of the horizontally-striped CO 
phase is a cooperative bond-charge density wave rather than a fully localized 
Wigner crystal.

\section{Sample characterization and experimental methods}
DC resistivity was measured between room temperature and 40~K by standard 
four-contact technique. In the frequency range 0.01~Hz--10~MHz the spectra of 
the complex dielectric function were obtained from the two-contact complex 
conductance measured by two set\-ups. The low-frequency, high-impedance setup 
covers the 0.01~Hz -- 3~kHz range. ac voltage signal is applied to the sample, 
the current response is transformed to voltage by a Stanford Research Systems 
SR570 current preamplifier and detected using a dual-channel digital lock-in 
Stanford Research Systems SR830. At higher frequencies (40~Hz -- 6~MHz) an 
Agilent 4294A impedance analyzer with virtual ground method was used. Even 
though the impedance analyzer reaches frequencies up to 110~MHz, we are limited to 
approximately 6~MHz by cable length. The applied ac signal levels, typically 50~mV 
(1~V/cm), were well within the linear response regime (upper bound verified 
to be at least 6~V/cm). The results obtained by both methods agree in the 
overlapping frequency range. In dielectric measurements a background 
contribution of stray capacitances is always present due to cabling and sample 
holder construction. In order to account for and remove these influences, we 
have routinely subtracted the open-circuit admittance from all measured sample 
admittances. The background capacitance of our setup amounts to 350~fF at all 
measured temperatures. At frequencies 10--10000~cm$^{-1}$ the complex dielectric 
function was obtained by a Kramers-Kronig analysis of the temperature-dependent 
infrared reflectivity measured by Fourier transform spectroscopy applying common 
procedure.\cite{Dressel04}

The samples under investigation are flat, planar high-quality single crystals. 
As a rule, the pronounced sample surface is in the $ab$ plane of the crystal 
structure. The $c$-axis of the crystal corresponds to the direction 
perpendicular to sample plane. The reflectivity measurements were performed on 
as-grown surfaces; for our $c$-axis investigation we employed an infrared 
microscope. 

Contacts for dc and ac conductivity measurements were made by applying carbon 
paint directly to sample surface. On our first studied single crystal contacts 
have been prepared parallel to sample edges without prior orienting the sample. 
After dc and ac conductivity measurements, the orientation of contacts was 
determined to be approximately along $[1 \bar{1} 0]$ (diagonal) direction by 
taking the x-ray back-reflection Laue photograph with the beam direction 
perpendicular to the largest facets of the sample. OrientExpress3.3 
software\cite{Milat} was applied to simulate the recorded flat back reflection 
patterns. Additional experiments were conducted on needle-like samples cut from 
one single crystal that was oriented by recording the mid-infrared spectra. The 
needles were cut along in-plane axes which ensured that the orientation of 
electric field is along one of the axes. The dc and ac conductivity on those 
samples was measured parallel to the $a$-axis (perpendicular to stripes; along 
this direction CR and CP sites alternate) and parallel to the $b$-axis (along 
the stripes).

With scrutiny we ruled out any influence of extrinsic effects in the dielectric 
spectroscopy measurements, especially those due to contact resistance and 
surface layer capacitance. Of the two relaxation modes detected in the 
dielectric response, the smaller one has a temperature-independent relaxation 
time, thus it can be identified as an intrinsic property of the sample. On the 
other hand, the larger dielectric mode bears features which might indicate a 
Maxwell-Wagner-like relaxation at the contacts\cite{Lunkenheimer02} and requires 
a thorough verification of contact quality. With this in mind, we performed dc 
resistance measurements in the standard four- and two-contact configurations. 
Taking into account the difference in geometry between contacts in these two 
configurations, the four-point resistance can be scaled and subtracted from the 
two-point resistance in order to estimate contact resistance ($R_\mathrm{c}$) 
\vs{}\ sample bulk resistance ($R_\mathrm{s}$). In the insulating phase our 
samples feature a gradually rising $R_\mathrm{c}/R_\mathrm{s}$ ratio, from 
0.1 just below the transition up to the order of 10 at temperatures below 50~K. 
However, independent of the changes in $R_\mathrm{c}/R_\mathrm{s}$ ratio, 
the strength of dielectric response remained approximately constant and finite 
throughout the insulating regime proving that the observed behavior of measured 
capacitance is governed predominantly by sample bulk response. Only the 
dielectric spectra in the immediate vicinity of $T_\mathrm{CO}$ were under 
influence of instrument artefact and therefore disregarded. Namely, above 135~K
the sample conductivity becomes greater than the imaginary part of admittance
$2\pi f C$ by a factor of 1000 or more in the whole range of frequency $f$,
an instrument limit beyond which no meaningful measurements of capacitance
$C$ can be made.

\section{Results and analysis}
\subsection{Optics}
Despite the numerous reports on the optical properties of 
\alphaETI{},\cite{Sugano85,Meneghetti86,Yakushi87,Yue10,Zamboni89,Zelzny90,
Gartner91,Moldenhauer93,Dressel94,Drichko09,Clauss09}
a few aspects of the development right at or below the charge-order transition 
are worthwhile to be reconsidered in the present context. Here we concentrate on 
two issues. The first is the conductivity and reflectivity in the 
highly-conducting ($ab$) plane at the metal-to-insulator phase transition which 
yield information on the anisotropy and the energy gap. The second issue 
concerns the redistribution of charge on the molecular sites that can be 
monitored via the vibrational features and their evolution on cooling. 
Vibrational spectroscopy with light polarized $\mathbf{E} \parallel c $ is most 
sensitive to characterize charge order.\cite{Dressel04,Yamamoto05} In addition, 
we present and discuss vibrational features of the BEDT-TTF molecule in metallic 
and the CO state seen with light polarized $\mathbf{E} \parallel a$ and $b$.

\begin{figure} 
\includegraphics[clip,width=1.0\columnwidth]{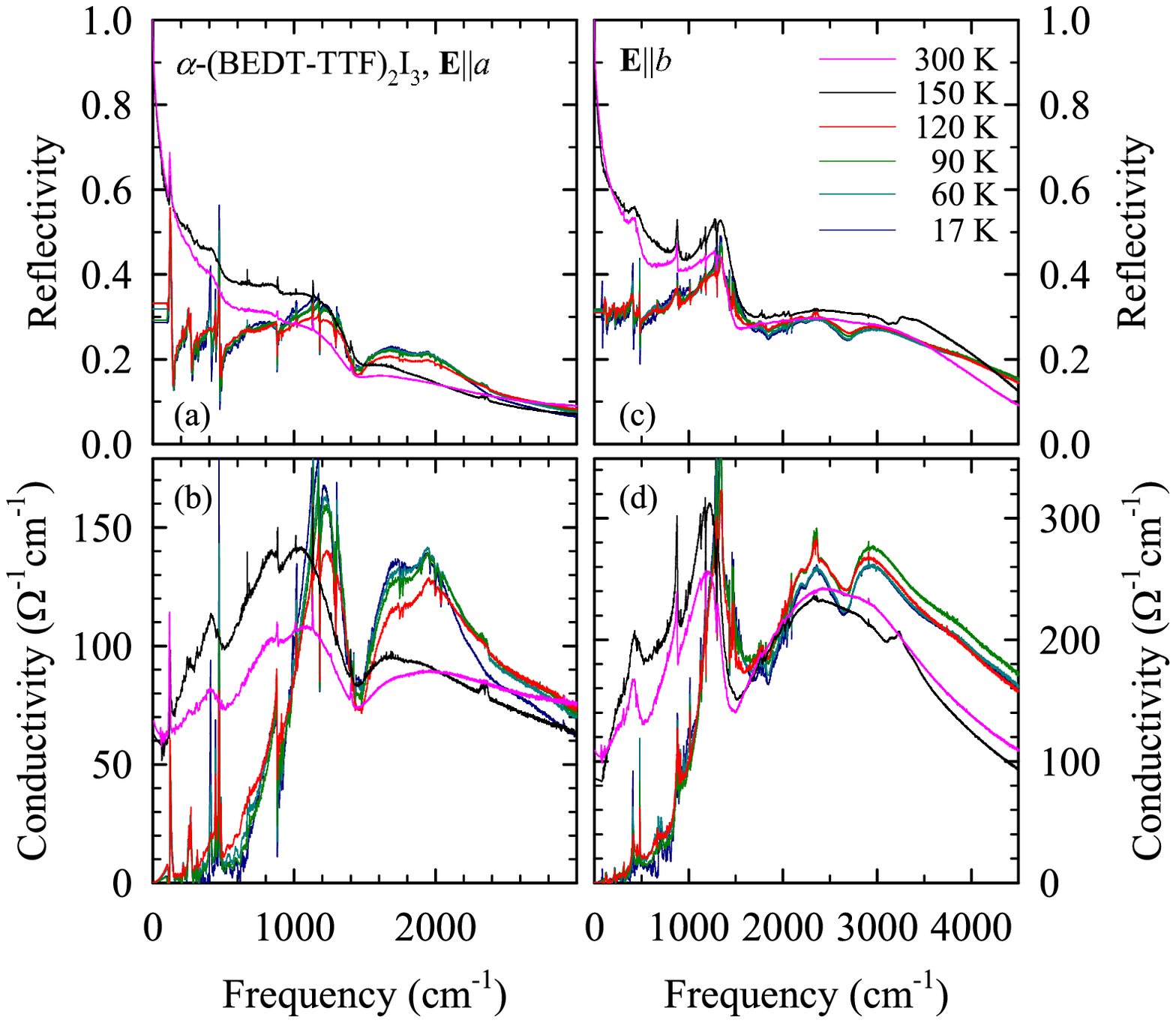}
\caption{\label{fig:refconab} (Color online) Optical properties of \alphaETI{}
for different temperatures as indicated. The upper panels (a) and (c) show the 
reflectivity, the lower panels (b) and (d) the corresponding conductivity. Note 
the different vertical scales. On the left side (a) and (b) measurements are 
shown with the electric field polarized parallel to the stacks 
($\mathbf{E}\parallel a$); the panels on the right side display the data for the 
polarization perpendicular to stacks ($\mathbf{E}\parallel b$).}
\end{figure}

\subsubsection{Electronic contributions}
The experimentally accessible frequency range extends from 10 to 5000~\cm{} and 
covers the bands formed by the overlapping orbitals of neighboring molecules. 
Fig.\ \ref{fig:refconab} shows the optical properties for the two polarizations 
$\mathbf{E} \parallel a$ and $\mathbf{E} \parallel b$ in the highly conducting
plane at different temperatures above and below $T_\mathrm{CO}$. The optical 
spectra are dominated by a broad band in the mid-infrared in both directions 
that is different in strength by about a factor of 2. A shoulder in $R(\omega)$, 
which shows up as a pronounced dip in the conductivity spectra around 1450~\cm, 
is due to the strong electron-molecular vibrational (emv) coupling of the
$\nu_{3}$(A$_g$) mode.\cite{Dressel04}

\begin{figure} 
\includegraphics[clip,width=0.7\columnwidth]{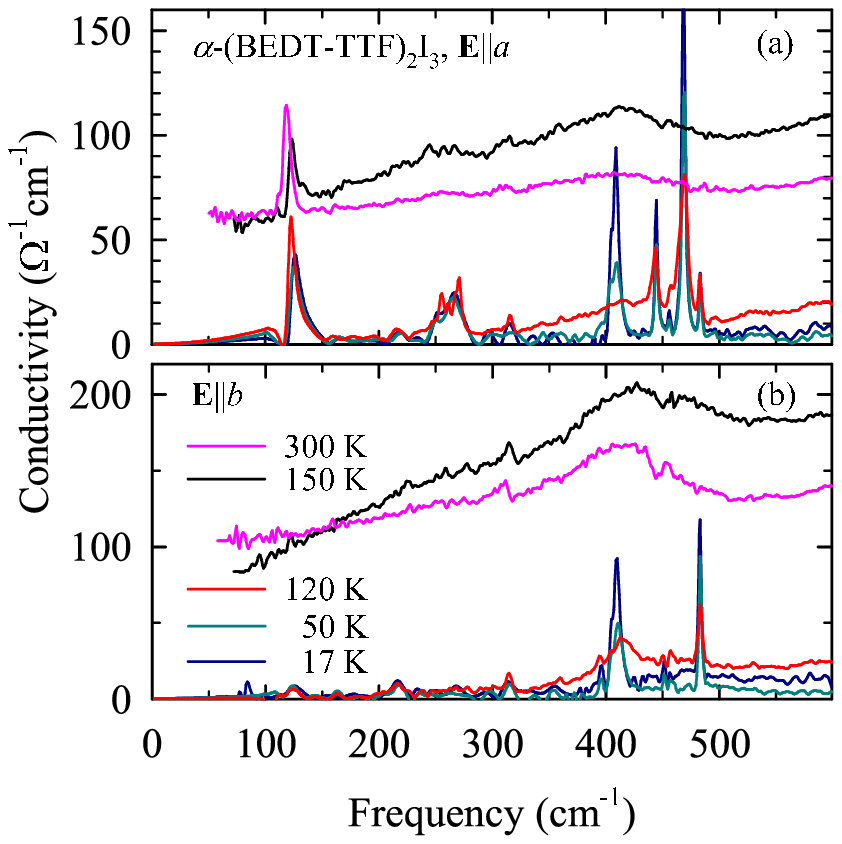}
\caption{\label{fig:condab} (Color online)  The far-infrared conductivity for
(a) $\mathbf{E} \parallel a$ and (b) $\mathbf{E} \parallel b$ for different
temperatures above and below the charge order transition at
$T_\mathrm{CO}=136$~K.}
\end{figure}

\paragraph{Overall behavior and anisotropy.}
Although the reflectivity exhibits a metallic response at ambient temperatures 
leading to a finite conductivity, no Drude-like response of the quasi-free 
carriers can be separated from the wing of the mid-infrared band
[Fig.\ \ref{fig:refconab}(b), (d) and Fig.\ \ref{fig:condab}]. Hence the weakly
temperature-dependent conductivity above the metal-insulator transition is 
described by an overdamped Drude response, \ie{}, a small spectral weight compared to 
the large scattering rate.

Upon decreasing the temperature from 300~K down to $T_\mathrm{CO}$, the 
reflectivity slightly increases due to reduced phonon scattering. In the CO 
state, the far-infrared reflectivity drops dramatically and the corresponding 
optical conductivity decreases as the energy gap opens in the density of states 
(Fig.\ \ref{fig:condab}). The spectral weight shifts to the mid-infrared range 
where it piles up in a band with maxima around 1500~\cm{} for 
$\mathbf{E}\parallel a$ (parallel to the stacks) and 2000~\cm{} for 
$\mathbf{E}\parallel b$ (perpendicular to the stacks).\cite{note2} When 
screening by the conducting charge carriers is reduced at $T<T_\mathrm{CO}$, the 
Fano-shaped antiresonances in the conductivity due the emv coupled molecular 
vibrations become even more pronounced and split the mid-infrared peaks.

\paragraph{Optical gap.}
\begin{figure} 
\includegraphics[clip,width=0.9\columnwidth]{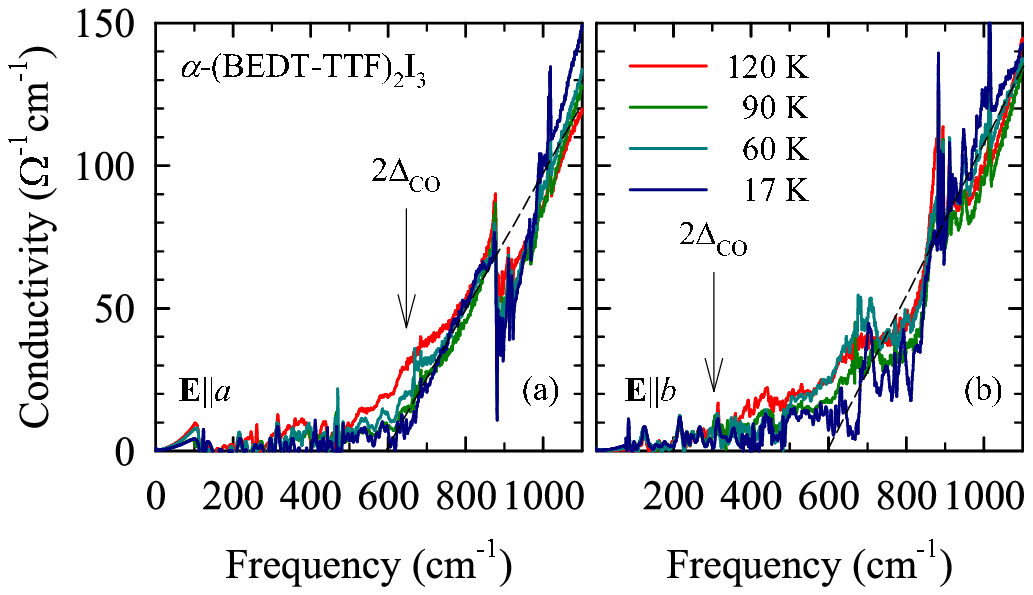}
\caption{\label{fig:gap} (Color online) Temperature dependence of the optical 
conductivity of $\alpha$-(BEDT-TTF)$_2$I$_3$ for $T<T_\mathrm{CO}$ measured for 
the polarization (a) $\mathbf{E} \parallel a$ and (b) $\mathbf{E} \parallel b$. 
In order to illustrate the development of the charge-order gap, the phonon lines 
have been subtracted to some extent. Dashed line shows the linear extrapolation 
which gives the optical gap value of about 600~cm$^{-1}$. Arrows denote the 
anisotropic dc transport gap.}
\end{figure}

At the metal-insulator phase transition, we see an abrupt opening of an optical 
gap, with the conductivity in the overdamped Drude region dropping down to very 
low values as shown in Fig.\ \ref{fig:condab}. In order to illustrate the
low-temperature electronic behavior more clearly, we have fitted the vibrational 
features by Lorentz and Fano curves, and subtracted them from the measured 
spectra. The results for both polarizations are plotted in Fig.\ \ref{fig:gap} 
for different temperatures. The drop of $\sigma(\omega)$ below 1000~\cm{} can be 
extrapolated linearly to obtain a gap value $2\Delta_0\approx 600$~\cm{} for 
$T\rightarrow 0$, corresponding to approximately 75~meV; it is basically 
identical for both polarizations.\cite{note3} It is worth of noting that while 
the conductivity for $\mathbf{E} \parallel a$ is indeed close to zero at 
frequencies below 600~\cm{}, the conductivity for $\mathbf{E} \parallel b$ 
remains finite down to about 400~\cm{}. Thus taking only the range up to 
800~\cm{} into account, we can extract gap values of 600 and 400~\cm{} from the 
linear extrapolation, which corresponds rather well to the one extracted from
the dc conductivity measurements (see Fig.\ \ref{fig:dcab}).

\begin{figure} 
\includegraphics[clip,width=0.9\columnwidth]{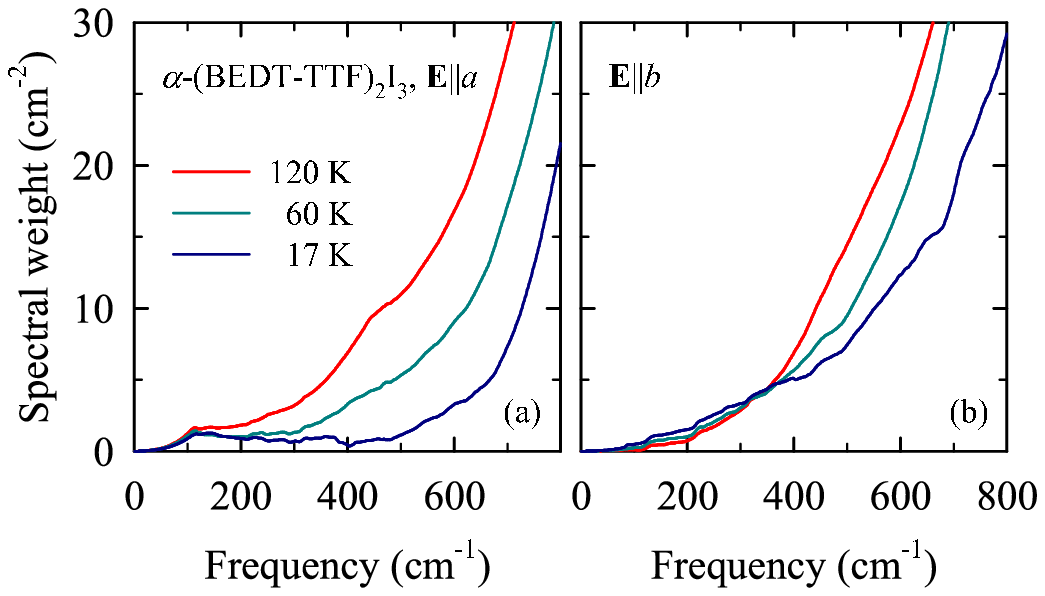}
\caption{\label{fig:sw} (Color online)  Development of the spectral weight 
SW$(\omega_c)$ as a function of cut-off frequency $\omega_c$, calculated by 
$\mathrm{SW}(\omega_c) = 8\int_0^{\omega_c}\sigma_1(\omega)\,\mathrm{d}\omega = {\omega_p^2}={4\pi n e^2}/{m}$
for both directions of \alphaETI{}.}
\end{figure}

For $T<T_\mathrm{CO}$ spectral weight still moves from the gap region to higher 
frequencies as $T$ is reduced: it piles up around 1000~\cm{} and higher
(Fig.\ \ref{fig:sw}). Interestingly, not only the region of the gap changes, but
spectral weight in the entire range shifts to higher frequencies. The maximum of
the mid-infrared band moves up slightly which in part can be described to
thermal contraction, but mainly to the redistribution of spectral weight.

\subsubsection{Vibrational features}
\paragraph{Charge disproportionation.}
\begin{figure} 
\includegraphics[clip,width=0.7\columnwidth]{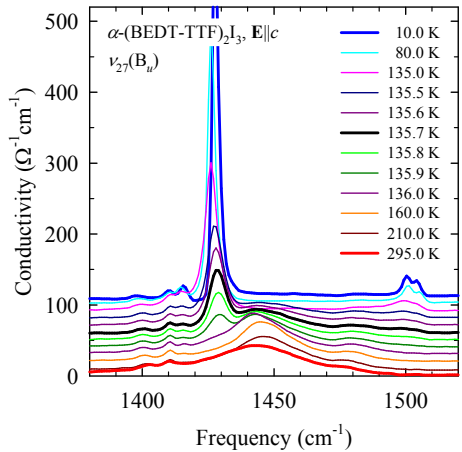}
\caption{\label{fig:phononsc} (Color online) Temperature dependence of the 
intramolecular vibrations of the BEDT-TTF molecule measured for the 
perpendicular direction $\mathbf{E} \parallel c$. The curves for different 
temperatures are shifted by 10~$(\Omega\mathrm{cm})^{-1}$ for clarity reasons. 
The $\nu_{27}$ mode becomes very strong right at the charge order transition 
($T=\mathbf{295}$, 210, 160, $\mathbf{136.0}$, 135.9, 135.8, $\mathbf{135.7}$, 
135.6, 135.5, $\mathbf{135.0}$, 80, and $\mathbf{10}$~K).}
\end{figure}

In order to characterize the redistribution of charge values on molecular sites 
associated with the charge ordering we followed the behavior of the infrared 
active $\nu_{27}(\mathrm{B}_u)$ charge-sensitive mode by measuring perpendicular 
to the conducting plane. This vibration is the out-of-phase contraction of the 
C=C double bonds in the BEDT-TTF rings which leads to a dipole-moment change 
parallel to the long axis of BEDT-TTF molecule. The frequency of this mode is 
sensitive to the charge population of the molecule and known to split upon 
passing through a charge-ordering phase
transition.\cite{Moldenhauer93,Yamamoto05} In the metallic state 
($T>T_\mathrm{CO}$) we observe a wide single band at about 1445~\cm{}; the 
frequency corresponds to an average charge of $+0.5e$ per molecule. The charge
diproportionation happens abruptly at $T_\mathrm{CO}=136$~K: in the CO
insulating state the mode splits in two pairs of bands at 1415 and 1428~\cm{},
and at 1500 and 1505~\cm{}, as demonstrated in the waterfall plot of Fig.\ \ref{fig:phononsc}.
The lower-frequency bands correspond to approximately $+0.8$ and $+0.85e$ charge
on the molecule, the upper-frequency modes to $+0.2$ and $+0.15e$. This charge 
redistribution remains constant on further cooling and is in agreement with the
charges estimated by x-ray for the four different sites in the unit cell.\cite{Kakiuchi07} 
The comprehensive infrared and Raman experiments of Yue \etal{}\cite{Yue10}
confirm our findings.

\begin{figure} 
\includegraphics[clip,width=0.9\columnwidth]{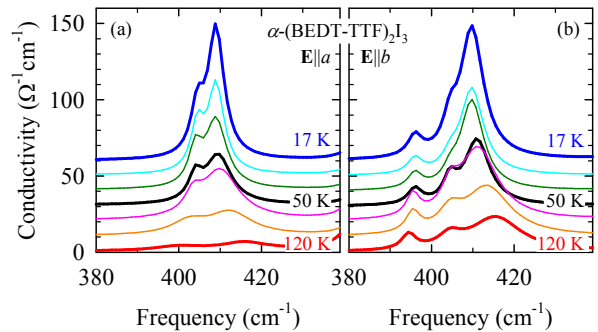}
\caption{\label{fig:phononsab} (Color online) Temperature dependence of the 
intramolecular vibrations $\nu_{14}$(A$_g$) of the BEDT-TTF molecule. The curves 
for different temperatures are shifted by 10~$(\Omega\mathrm{cm})^{-1}$ for 
clarity reasons. For the polarizations (a) $\mathbf{E} \parallel a$ and (b) 
$\mathbf{E} \parallel b$ the $\nu_{14}$ grows and is splits in three distinct 
peaks ($T=\mathbf{120}$, 90, 60, $\mathbf{50}$, 40, 30, and $\mathbf{17}$~K).}
\end{figure}

Coming back to the vibrational features; also for $\mathbf{E}$ polarized 
parallel $a$ and $b$, we observe changes in shape of some modes at the 
metal-insulator transition. Due to screening, no molecular vibrations can be 
seen above the metal-to-insulator phase transition. Below $T_\mathrm{CO}$, the 
modes detected in the in-plane spectra are the features of totally symmetric 
vibrations of BEDT-TTF molecule emv coupled with electronic charge-transfer 
transitions which have been observed and assigned previously\cite{Meneghetti86} 
down to 300~\cm{}. As a result of interaction with electronic transition they 
have a Fano-shape:\cite{Fano61} an anti-resonance at frequencies where they 
coincide with the electronic excitations and an asymmetric peaks shape when the 
electronic feature is separated in frequencies. Thus, for example the 
A$_g(\nu_3)$ feature and several symmetric and asymmetric CH$_3$ 
vibrations\cite{Meneghetti86} at about 1400~\cm{} not only shows a blue shift 
together with a charge-transfer band in the mid-infrared, but also changes shape 
to become a narrow and slightly asymmetric band (see Fig.\ \ref{fig:refconab}). 
While the lower-frequency modes are only weakly seen for $T>T_\mathrm{CO}$, in 
the insulating phase the spectra we observe (cf.\ Fig.\ \ref{fig:condab}) all of 
the A$_g$ vibrations predicted by Meneghetti \etal{}:\cite{Meneghetti86} for 
instance, the $\nu_{15}(\mathrm{A}_g)$ mode at 260~\cm{} (associated with the 
deformation of the outer EDT rings), the $\nu_{16}(\mathrm{A}_g)$ mode at 
124~\cm{} (associated with the deformation of the inner TTF rings). These bands 
are very intense only in the $\mathbf{E} \parallel a$ direction and barely seen 
in the $b$ polarization. This is in agreement with the symmetry breaking, \ie{}, 
both the intrinsic dimerisation along the stacks ($a$ direction) and the stripes 
formed along the $b$ direction in the CO phase as depicted in
Fig.\ \ref{fig:a-ET-CO}(b). A more detailed discussion of the in-plane vibrational 
features was recently given by Yue \etal{} based on low-temperature
transmission experiments.\cite{Yue10}

Interesting, that only the 410~\cm{} mode plotted in Fig.\ \ref{fig:phononsab} 
changes on cooling in the insulating state. The band is much wider than the 
other features in this range at temperatures right below the metal-insulator 
transition and continuously narrows as $T$ is reduced. Following Meneghetti
\etal{}\cite{Meneghetti86} we assign it to the $\nu_{14}(\mathrm{A}_g)$
mode which mainly involves the deformation of the outer rings. 

Finally, the strong vibrational feature observed around 1300~\cm{} (not shown)
is assigned to the emv coupled $\nu_4(\mathrm{A}_g)$ mode of the BEDT-TTF
molecule. It is sharper and more pronounced for $\mathbf{E}\parallel a$ although
the overall conductivity in the mid-infrared is about half compared to
$\mathbf{E}\parallel b$. Below 1000~\cm{} a large number of molecular and
lattice vibrations peak out as soon as the screening by the conduction electrons
is lost.

\subsection{Transport}
\begin{figure} 
\includegraphics[clip,width=0.7\columnwidth]{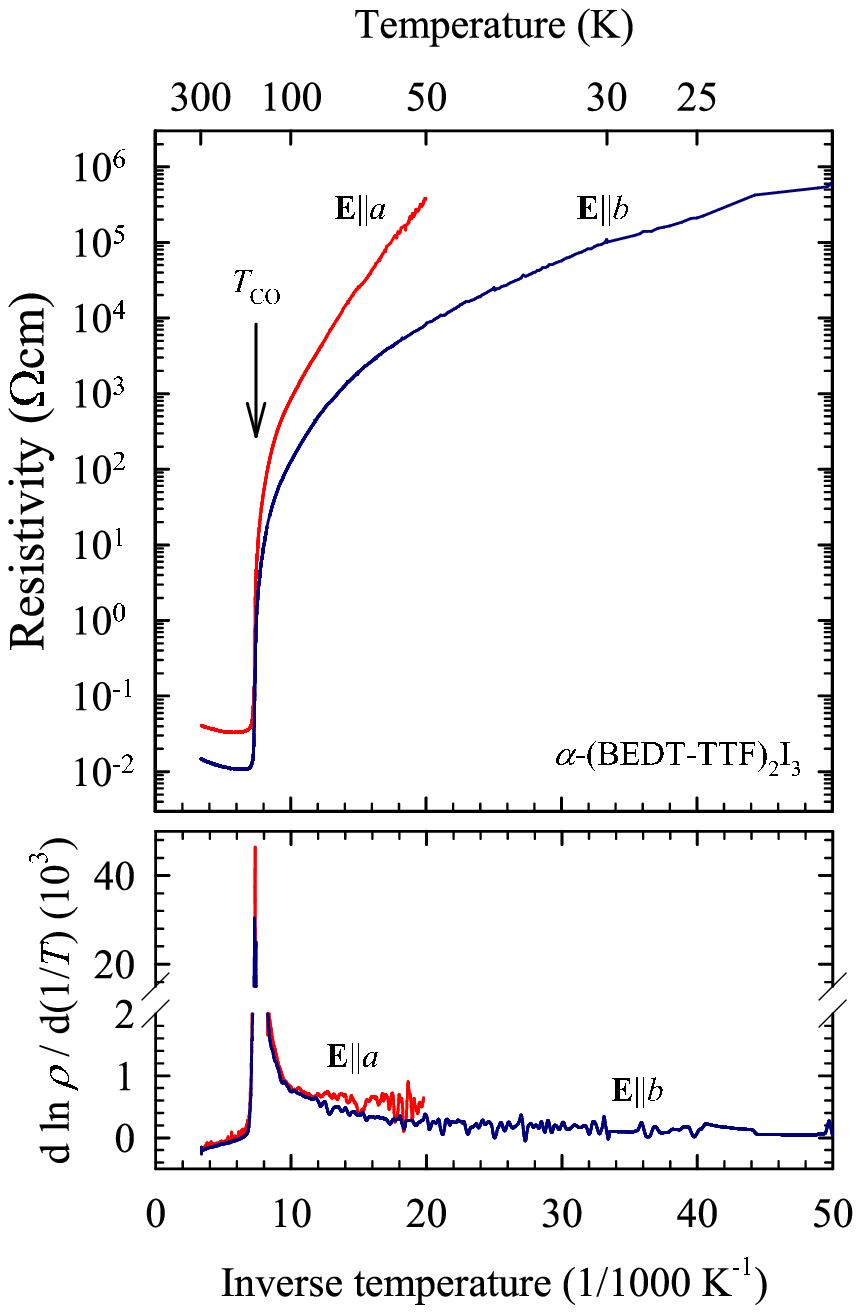}
\caption{\label{fig:dcab} (Color online) Resistivity (upper panel) and
logarithmic resistivity derivative (lower panel) \vs{}\ inverse temperature of
\alphaETI{} for $\mathbf{E}\parallel a$ (red line) and $\mathbf{E}\parallel b$
(blue line).}
\end{figure}

These measurements reveal that the small dc resistivity anisotropy known to be 
present at room temperature,\cite{Bender84-108} $\rho_a/\rho_b \approx 2$, 
pertains to the whole metallic regime and is approximately constant down to 
$T_\mathrm{CO}$. As a new result it has to be pointed out that below 
$T_\mathrm{CO}$ the anisotropy of resistivity, hence also of conductivity, 
changes significantly with lowering temperature as the resistivity along the 
$a$-axis rises more steeply than along the $b$-axis, and at 50~K reaches 
$\rho_a/\rho_b = 50$ (see Fig.\ \ref{fig:dcab}). In our samples, despite 
temperature-dependent activation, the anisotropic transport gap in the CO phase 
for $\mathbf{E}||a$ and $\mathbf{E}||b$ can be estimated to about
$2\Delta = 80$~meV and 40~meV, respectively. At a first glance the appearance of
an anisotropic transport gap seems to be at odds with the isotropic optical gap 
extracted from our optical measurements (Fig.\ \ref{fig:gap}). It is well known 
that systems with a complex band structure such as \alphaETI{} may exhibit quite 
different optical and transport gaps: optical measurements examines direct 
transitions between the valence and conduction band, while dc transport probes 
transitions with the smallest energy
difference between the two bands.

\begin{figure} 
\includegraphics[clip,width=0.7\columnwidth]{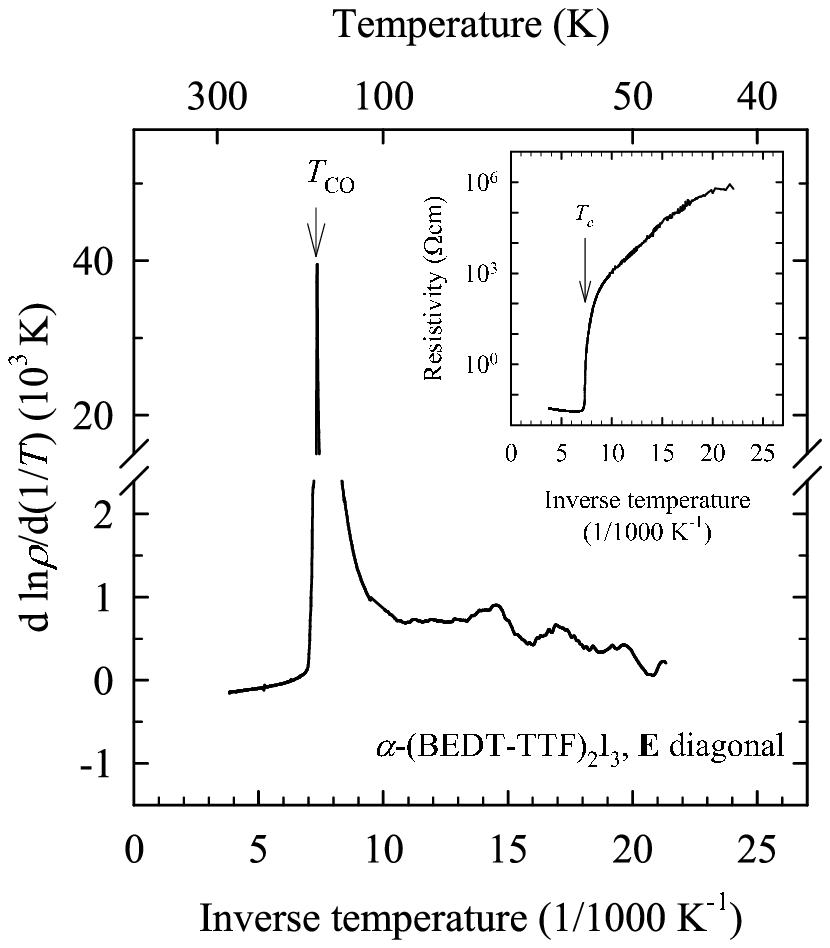}
\caption{\label{fig:dcdiag}Logarithmic resistivity derivative (main panel) and 
resistivity (inset) \vs{}\ inverse temperature of \alphaETI{} for 
$\mathbf{E}\parallel [1 \bar{1} 0]$, \ie{}, in the diagonal direction of the
$ab$ plane.}
\end{figure}

We have also characterized dc resistivity of \alphaETI{} in the conducting $ab$ 
plane, for $\mathbf{E}\parallel [1 \bar{1} 0]$, \ie{}, at an angle of 
approximately $45^\circ$ to the crystallographic axes. Metallic behavior of 
resistivity is present from room temperature down to 156~K. A sharp
metal-to-insulator transition is confirmed\cite{Dressel94} at
$T_\mathrm{CO} = 136.2$~K, which is apparent in the peak in
$\mathrm{d}(\ln\rho)/\mathrm{d}(1/T)$ with full width at half-height
$2\delta T_\mathrm{CO} =1.5$~K; $2\delta T_\mathrm{CO}/ T_\mathrm{CO} = 0.011$
(Fig.\ \ref{fig:dcdiag}). Below the transition the resistivity curve rises with
a temperature-dependent activation indicating that a temperature-dependent
conductivity gap opens of about 80~meV. No significant hysteresis in dc
resistivity in the vicinity of $T_\mathrm{CO}$ could be found.

\subsection{Dielectric Response}
\begin{figure} 
\includegraphics[clip,width=0.7\columnwidth]{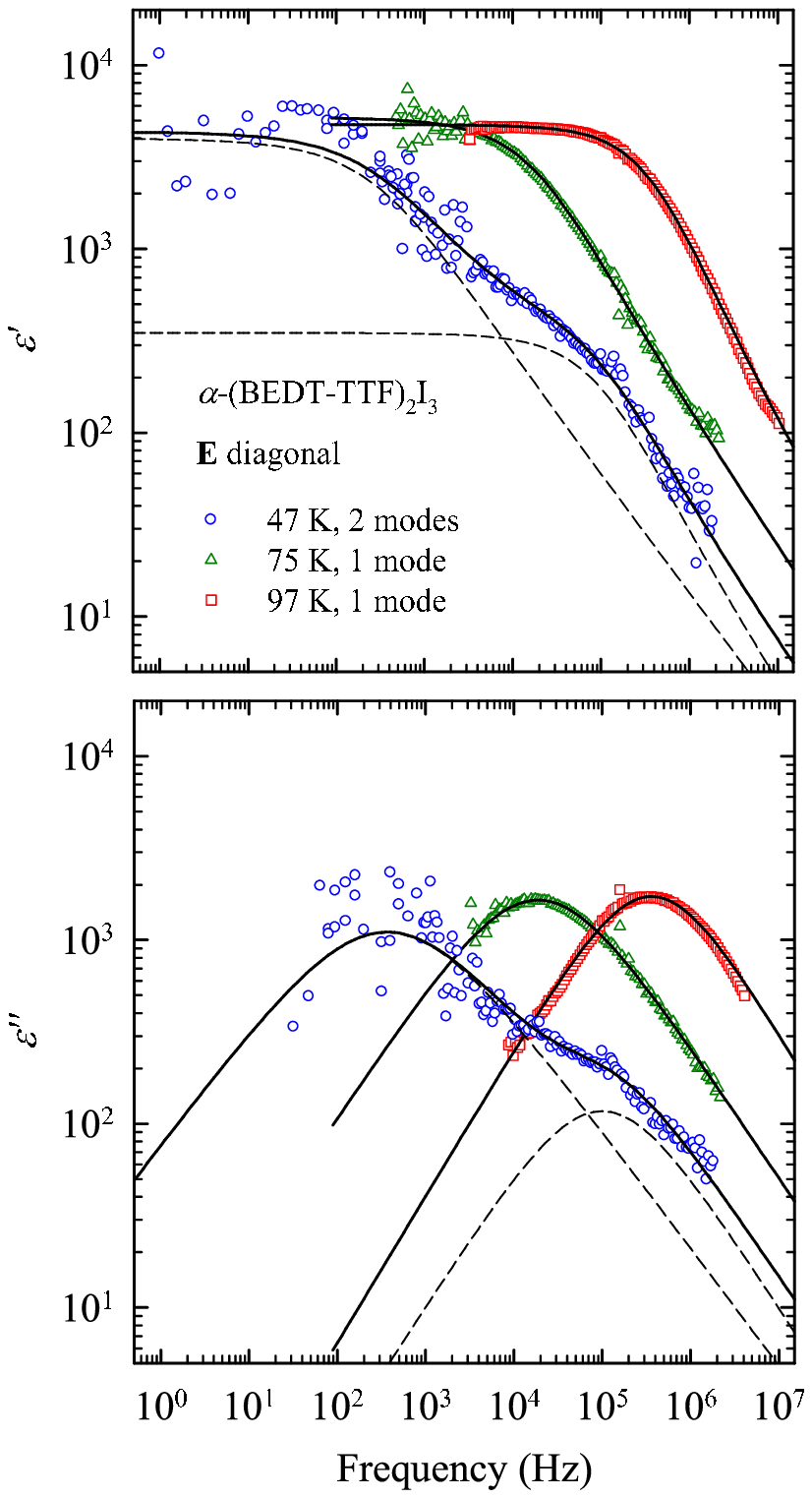}
\caption{\label{fig:lfdsspectra} (Color online) Double logarithmic plot of the frequency 
dependence of the real ($\varepsilon'$) and imaginary ($\varepsilon''$) part of 
the dielectric function in \alphaETI{} at representative temperatures for 
$\mathbf{E}\parallel [1\bar{1}0]$. Below 75~K two dielectric relaxation modes 
are observed -- full lines for 47~K show a fit to a sum of two generalized Debye 
functions; dashed lines represent contributions of the two modes. Above 75~K 
only one mode is detected, and the full lines represent fits to single 
generalized Debye functions.}
\end{figure}

\begin{figure} 
\includegraphics[clip,width=0.8\columnwidth]{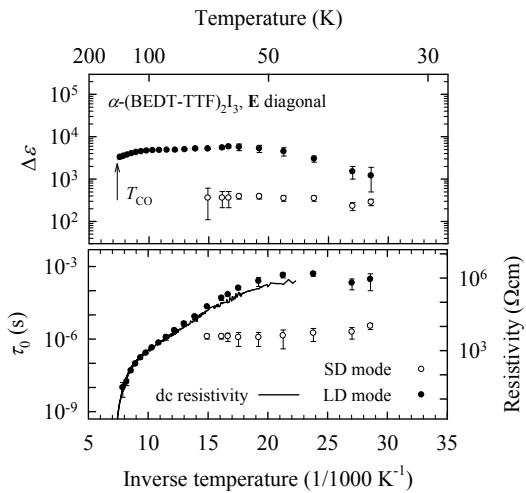}
\caption{\label{fig:lfdsparamsdiag}Dielectric strength (upper panel) and mean
relaxation time with dc resistivity (points and line, respectively, lower panel)
in \alphaETI{} as a function of inverse temperature, for
$\mathbf{E}\parallel [1\bar{1} 0]$.}
\end{figure}

Low-frequency dielectric spectroscopy measurements were performed at various 
temperatures in the semiconducting phase. Representative spectra for 
$\mathbf{E}\parallel [1 \bar{1} 0]$ are shown in Fig.\ \ref{fig:lfdsspectra}. 
Most notably, between 35~K and up to 75~K two dielectric relaxation modes are 
discerned. The complex dielectric spectra $\varepsilon(\omega)$ can be described 
by the sum of two generalized Debye functions
\begin{equation}
\varepsilon(\omega)-\varepsilon_\infty
 = \frac{\Delta\varepsilon_\mathrm{LD}}{ 1 + \left(i \omega \tau_{0,\mathrm{LD}} \right)^{ 1-\alpha_\mathrm{LD} } }
 + \frac{\Delta\varepsilon_\mathrm{SD}}{1 + \left(i \omega \tau_{0,\mathrm{SD}} \right)^{ 1-\alpha_\mathrm{SD} } }
\label{eqmodel}
\end{equation}
where $\varepsilon_\infty$ is the high-frequency dielectric constant, 
$\Delta\varepsilon$ is the dielectric strength, $\tau_0$ the mean relaxation 
time and $1-\alpha$ the symmetric broadening of the relaxation time distribution 
function of the large (LD) and small (SD) dielectric mode. The broadening 
parameter $1-\alpha$ of both modes is typically $0.70 \pm 0.05$. The temperature 
dependences of dielectric strengths and mean relaxation times are shown in
Fig.\ \ref{fig:lfdsparamsdiag}. The dielectric strength of both modes does not 
change significantly with temperature ($\Delta\varepsilon_\mathrm{LD} \approx 5000$,
$\Delta\varepsilon_\mathrm{SD} \approx 400$). At approximately 75~K the large 
dielectric mode overlaps the small mode. It is not clear whether the small 
dielectric mode disappears at this temperature or is merely obscured by the 
large dielectric mode due to its relative size. However, above 100~K, when the 
large dielectric mode shifts to sufficiently high frequencies, there are no 
indications of a smaller mode centered in the range $10^5$--$10^6$~Hz. 
Accordingly, above 75~K fits to only a single Debye function are performed that 
we identify with the continuation of the large dielectric mode. All parameters 
of the large mode -- such as dielectric strength, mean relaxation time, 
symmetric broadening of the relaxation time distribution function -- can be 
extracted in full detail until it exits our frequency window at approximately 
130~K. At temperatures up to 135~K (just below $T_\mathrm{CO} = 136$~K) we can 
determine only the dielectric relaxation strength by measuring the capacitance 
at 1~MHz.

One of the most intriguing results is that the temperature behavior of the mean 
relaxation time differs greatly between the two dielectric modes. The large 
dielectric mode follows a thermally activated behavior similar to the dc 
resistivity, whereas the small dielectric mode is almost
temperature-independent. This unexpected and novel behavior in the
charge-ordered phase raised the possibility of anisotropic dielectric response.
With this in mind we have performed another set of dc and ac spectroscopy
measurements on the needle-shaped samples oriented along the $a$- and $b$-axis.

\begin{figure} 
\includegraphics[clip,width=0.7\columnwidth]{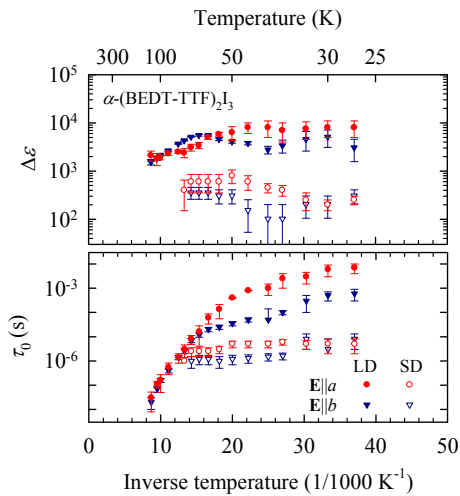}
\caption{\label{fig:lfdsparamsab} (Color online) Dielectric strength (upper panel)
and mean relaxation time (lower panel) in \alphaETI{} as a function of inverse 
temperature; full and empty symbols represent parameters of the large and small 
dielectric mode, respectively, for $\mathbf{E}$ along the $a$- (red circles) and 
$b$-axis (blue triangles). Compared to Fig.\ \ref{fig:lfdsparamsdiag}, the
relatively large error bars are due to a somewhat unfavorable sample geometry
which results in higher resistances and a smaller capacitive response.}
\end{figure}

\begin{figure} 
\includegraphics[clip,width=0.7\columnwidth]{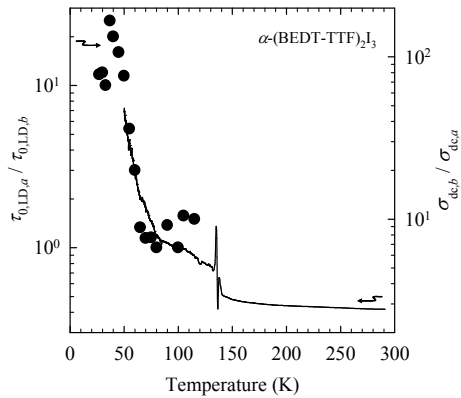}
\caption{\label{fig:anisotropy}Anisotropy of the large-dielectric-mode
mean-relaxation time (points) closely follows the temperature behavior of dc 
conductivity anisotropy (line) in \alphaETI{}.}
\end{figure}

Low-frequency dielectric spectroscopy for both $\mathbf{E}||a$ and 
$\mathbf{E}||b$ orientation yields results comparable to
$\mathbf{E}\parallel [1 \bar{1} 0]$: a large mode whose mean relaxation time 
follows dc resistivity, and a small, temperature-independent mode noticeable at 
temperatures below $T\approx 75$~K. The fit parameters to model (\ref{eqmodel}) 
are displayed in Fig.\ \ref{fig:lfdsparamsab} as a function of inverse 
temperature. There is no prominent anisotropy or temperature dependence in 
dielectric strength, and the $\Delta\varepsilon$ values of both the large and 
small dielectric modes correspond to those of the sample measured in 
$\mathbf{E}$ diagonal orientation. However, an evolution of anisotropy in 
$\tau_{0,\mathrm{LD}}$ is clearly visible. Figure \ref{fig:anisotropy} shows 
that the newly-found anisotropy in $\tau_{0,\mathrm{LD}}$ closely follows the dc
conductivity anisotropy.

\section{Discussion}
\begin{figure} 
\includegraphics[clip,width=0.7\columnwidth]{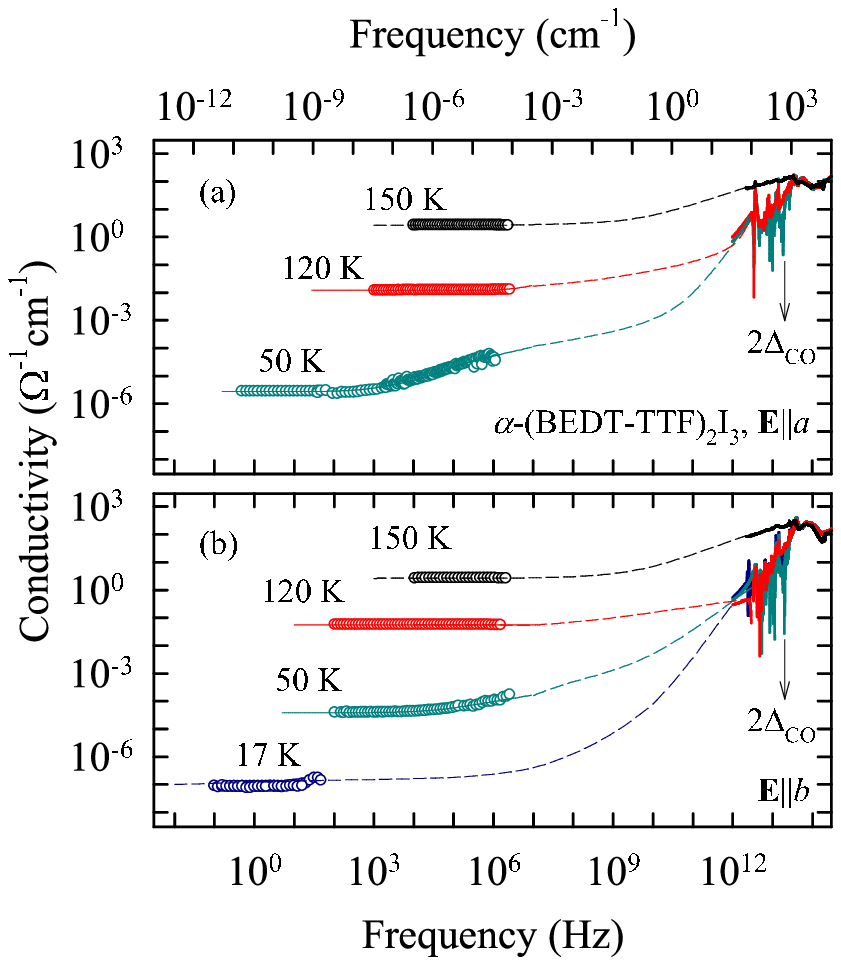}
\caption{\label{fig:broadband} (Color online) Broad-band conductivity spectra of \alphaETI{}
for (a) $\mathbf{E} \parallel a$ and (b) parallel $b$ at a few selected 
temperatures. Vertical arrows show the CO optical gap. The dashed lines are
guides for the eye.} 
\end{figure}

It is instructive to compare the conductivity of \alphaETI{} as a function of 
temperature in the wide frequency range: from dc limit up to terahertz 
frequencies. Fig.\ \ref{fig:broadband} composes the conductivity spectra of 
\alphaETI{} from dc, dielectric and optical measurements for $\mathbf{E} 
\parallel a$ and $b$ at different temperatures. First, we address the 
high-temperature phase. The Drude term in room-temperature spectra of organic
conductors is known to be commonly very weak, if present at all.\cite{Dressel04}
The optical conductivity of \alphaETI{} is in accord to this observation, 
showing both along
the $a$- and $b$-axes an overdamped Drude response at all temperatures above 
$T_\mathrm{CO}$. The absence of a well-defined
Drude peak in the vicinity of the CO transition resembles the behavior
reported for the CO insulator $\theta$-(BEDT-TTF)$_2$RbZn(SCN)$_4$.\cite{Wang01}
Note that other $\alpha$-phase BEDT-TTF conductors, such as 
$\alpha$-(BEDT-TTF)$_2M$Hg(SCN)$_4$,\cite{Drichko06} do exhibit a zero-energy 
peak.

It is also worth of noting that the optical, dc resistivity and low-frequency 
dielectric measurements give mutually consistent values for the conductivity 
anisotropy $\sigma_b/\sigma_a\approx 2$ at all temperatures above 
$T_\mathrm{CO}$. The electronic part of optical spectra can be compared to 
calculations of the extended Hubbard model for a quarter-filled square lattice 
using Lanczos diagonalization. They predict a band with a maximum at 
approximately $6t$ in the charge ordered insulating state,\cite{Merino03} which 
yields $t_a= 0.03$~eV and $t_b=0.04$~eV for the respective directions. The 
values are in reasonable agreement with H\"{u}ckel calculations performed by 
Mori \etal{}\cite{MoriCL84} and support the observed anisotropy in transport and 
optical properties.

Next we address the charge-order phase: how it develops on cooling, the ground 
state features and excitations observed by applied spectroscopic techniques. The 
vibrational spectra reveal that the static charge disproportionation sets in 
rather suddenly (Fig.\ \ref{fig:phononsc}) at the temperature of MI transition 
and is accompanied by the respective change in the optical properties of the 
conducting plane. At high temperatures, a wide single band at about 1445~\cm{} 
is observed whose frequency corresponds to an average charge of $+0.5e$ per 
molecule. According to Yue \etal{}\cite{Yue10} this band originates from slow 
fluctuations of the charge distribution at each site reflecting the partial 
charge ordering at short length scales as detected in NMR and x-ray 
measurements.\cite{Takahashi06,Kakiuchi07} Yue \etal{} estimate the site-charge
distribution slightly above transition to be +0.6, +0.6 and $+0.4e$, agreeing
well with the x-ray data by Kakiuchi \etal{} The long-range charge
diproportionation happens abruptly at $T_\mathrm{CO}=136$~K and remains constant 
on further cooling. In the CO insulating state the mode splits in two pairs of
bands (see Fig.\ \ref{fig:phononsc}). The lower-frequency bands correspond to
approximately $+0.8$ and $+0.85e$ charge on the molecule, and the
upper-frequency modes to $+0.2$ and $+0.15e$, which is in agreement with charge
estimation by x-ray for the four different sites in the unit cell.\cite{Kakiuchi07} 

Interestingly though, optical gap and some of the features of outer ring 
BEDT-TTF vibrations show a continuous change on cooling in the CO state, 
indicating that some changes (but not a charge redistribution) happen with 
temperature in the insulating state. In contrast to the sharp onset of the 
$\nu_{27}(\mathrm{B}_u)$ vibration that monitors the static charge order and 
does not change below $T_\mathrm{CO}$ (see Fig.\ \ref{fig:phononsc}), we see 
some development of the gap and some of the emv coupled features on cooling in 
the charge-ordered state. In the region where the gap has opened 
($T<T_\mathrm{CO}$), the conductivity drops further and reaches zero only at the 
lowest measured temperature ($T=17$~K). For instance, at $T=120$~K a finite 
conductivity is found all the way down to 200~\cm{} and even below, in accord 
with previous microwave measurements.\cite{Dressel94,Clauss09} The optical gap
is more or less isotropic, in contrast to the pronounced anisotropy of the dc
gap, which is explainable taking into account that different transitions are
involved in optics and dc. Nevertheless, as mentioned in Section III, there is
some weak indication that for $\mathbf{E}\parallel b$ excitations are possible
to lower frequencies. The increase of the anisotropy at lower temperatures,
which was observed in the dc limit, is not that clear in the optical data
possibly due to strong phonon features and the low-conductivity base line.
A similar dc conductivity anisotropy has been observed in the CO phase of
(TMTTF)$_2$AsF$_6$.\cite{KorinHamzic06}

In the charge-ordered phase we observe novel processes at lower frequencies, 
including CDW responses. As soon as the CO phase is entered, the low-frequency 
conductivity drops strongly leading to a step in the radio-frequency range
(see Fig.\ \ref{fig:broadband}). This corresponds to the broad and strongly 
temperature-dependent CDW relaxation mode (visible only below 120~K), which can 
be clearly seen in the spectra of imaginary part of dielectric function
(Fig.\ \ref{fig:lfdsspectra}). It is followed by a power law dispersion
attributed to hopping transport (for more details on hopping conduction see
Ref.\ \onlinecite{VuleticPR}), that leads to relatively high ac conductivity in
the microwave and far-infrared region, as compared to the conductivity in dc
limit (see Fig.\ \ref{fig:broadband}). In the microwave region, the most
prominent feature is the continuous increase of conductivity with rising
frequency, while the far-infrared and infrared regions are mainly characterized
by the suppression of the Drude weight, below either the CO gap and strong
phonon features. Such behavior of conductivity is similar to the one observed
for fully doped ladders in the (Sr,Ca)$_{14}$Cu$_{24}$O$_{41}$ cuprates in which
CDW is established. Conversely, comparable dc and optical conductivities were
found in BaVS$_3$ systems in which CDW is also observed.\cite{Kezsmarki06} In
this way, we can classify the latter system as the fully ordered, while ladders
and \alphaETI{} show features known for disordered systems.

\begin{figure} 
\includegraphics[clip,width=0.9\columnwidth]{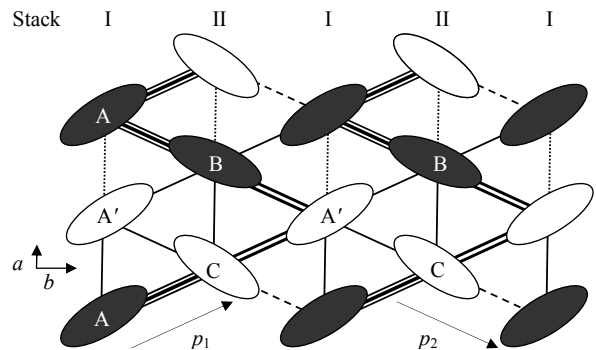}
\caption{\label{fig:bonds}Schematic representation of a $2k_\mathrm{F}$ bond-CDW 
in \alphaETI{}. Triple, double, single and dashed lines show relative strengths 
of overlap integrals, from strongest to weakest, along $p_1$- and
$p_2$-directions\cite{Kakiuchi07}. Also denoted are dimerized bonds along the
$a$-axis in the AA$^\prime$ and BC stacks.}
\end{figure}

The observed ac conductivity data demonstrate a complex and anisotropic 
dispersion in the charge-ordered state. First, similar to the Peierls CDW state, 
we observe broad screened relaxation (large dielectric) modes along diagonal and 
both $a$- and $b$-axis of BEDT-TTF planes. These modes can be interpreted as 
signatures of long-wavelength charge excitations possessing an anisotropic 
phason-like dispersion. In \alphaETI{} Kakiuchi \etal{}\ were the first to 
suggest a $2k_\mathrm{F}$ CDW which forms along the zig-zag path CABA$^\prime$C 
of large overlap integrals detected in their x-ray diffraction measurements 
\cite{Kakiuchi07}. However, the presence of a $2k_\mathrm{F}$ modulation of 
overlap integrals along the $p_1$- and $p_2$-directions, ACA$^\prime$BA and 
ABA$^\prime$CA (see Fig.\ \ref{fig:bonds}), hints at an additional complexity 
and makes the zig-zag paths a somewhat arbitrary choice. A theoretical model for 
the related quarter-filled $\theta$-ET$_2X$ systems may provide a more 
appropriate scenario.\cite{Clay02} Clay \etal{} showed that the CO phase with 
horizontal stripe phase is characterized by a 1100 modulation of site charges 
along the two independent $p_1$- and $p_2$-directions parallel to the larger
overlap integral, and with a 1010 modulation along the $b$-axis perpendicular to
the stripes. In addition, this CO is accompanied by a $2k_\mathrm{F}$ modulation,
or tetramerization, of the overlap integrals along the $p$-directions, the 
strongest overlap integral being 1-1 and the weakest 0-0. Further, a bond 
dimerization is also present along the molecular stacks. In other words, such a 
CO phase corresponds to a specific modulation of bonds and site charges, \ie{}, 
a combined $2k_\mathrm{F}$ bond-CDW along the two BEDT-TTF plane $p$-directions 
with bond dimerization in stacking direction. As mentioned above, an analogous 
albeit more complex tetramerization of overlap integrals does develop along the 
$p$-directions of \alphaETI{}. Using x-ray diffraction data, Kakiuchi
\etal{}\ calculated overlap integrals between neighboring molecules based on the 
tight-binding approximation and a molecular orbital calculation with the 
extended H\"{u}ckel method.\cite{Kakiuchi07,MoriBCSJ84} As shown in
Fig.\ \ref{fig:bonds}, along the $p_2$-direction, ABA$^\prime$CA, the strongest 
overlap integral is obtained between the two charge-rich sites A and B, quite 
alike the model bond order for the $\theta$-material. Also, bond dimerization 
along the stacking $b$-direction and its pattern in the $ab$ plane of 
\alphaETI{} agree with the model. However, in the $p_1$-direction, 
ACA$^\prime$BA, the order is shifted by one bond: the largest overlap integral 
is between the charge-rich A site and the charge-poor C site. Additionally, the 
overlap integrals are not perfectly $2k_\mathrm{F}$ sine-modulated along each 
$p$-direction. While these deviations of \alphaETI{} bond order from the 
$\theta$-ET$_2X$ model should be recognized, they are hardly surprising. Indeed, 
\alphaETI{} has a lower symmetry than a $\theta$-structure which might induce 
slight differences in bond patterns. Also, the overlap integrals obtained from 
x-ray diffraction could somewhat depend on the employed method of calculation. 
This leaves the main physical result of the model by Clay \etal{}, formation of 
a bond-CDW within the conducting molecular planes, fully applicable and relevant 
to the case of \alphaETI{}. It is plausible to look for the origin of 
phason-like dielectric relaxation in such a $2k_\mathrm{F}$ bond-CDW. In this 
case the energy scale of barrier heights is close to the single-particle 
activation energy indicating that screening by single carriers responsible for 
the dc transport is effective for this relaxation. The fact that the temperature 
behavior of the $\tau_{0,\mathrm{LD}}$ anisotropy closely follows the dc 
conductivity anisotropy has important implications: while the CDW motion is 
responsible for the dielectric response, the single electron/hole motion along 
the two $p$-directions, possibly zig-zagging between them, is responsible for 
the observed dc charge transport.

The observation of a Peierls-like broad screened dielectric relaxation in 2D 
represents an important experimental result which clearly indicates that the 
charge order in \alphaETI{} cannot be considered of fully localized Wigner type 
as predicted by a number of theoretical models.\cite{Kino9596,Seo06} Rather, as 
we have argued above, the bond-CDW delocalized picture appears as the most 
appropriate one. It is noteworthy that a similar 2D dispersion was previously 
observed in the CDW phase developed in the ladder layers of 
Sr$_{14}$Cu$_{24}$O$_{41}$.\cite{VuleticPRL,Vuletic05,Abbamonte04} Calculations 
based on the extended Hubbard model and the $t$-$J$ model predicted 
sinusoidal charge-density modulation either of $2k_\mathrm{F}$ or 
$4k_\mathrm{F}$ type for this system.\cite{Dagotto92,Orignac97,White02} This 
result was at first met with surprise due to the strongly correlated nature of
the ladders system. Nevertheless, the prediction was confirmed by resonant x-ray 
diffraction measurements\cite{Abbamonte04} which indicated that the established 
CO is not of fully localized Wigner type, but that it can be described as a 
sinusoidal, delocalized modulation.
 
\begin{figure} 
\includegraphics[clip,width=0.7\columnwidth]{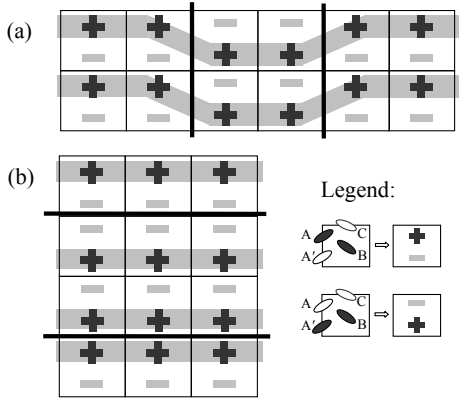}
\caption{\label{fig:domainwalls}Two different types of domain wall pairs in the 
charge-ordered phase of \alphaETI{}. (A,B)- and (A$^\prime$,B)-rich unit cells 
are symbolically represented as $+-$ or $-+$ cells which form CO stripes. For 
simplicity we omit the B and C molecules. Gray thick lines stand for charge-rich 
stripes. Thin black lines denote a domain wall pair.}
\end{figure}

Further, we comment on the small dielectric mode whose features are 
characteristic of short-wavelength charge excitations. We suggest that the 
origin of this relaxation might be in  the twinned nature of the CO phase due to 
the inversion symmetry breaking, with one domain being (A,B)-rich and the other 
(A$^\prime$,B)-rich.\cite{Kakiuchi07} Indeed, a ferroelectric-like character to 
the charge-ordered phase is suggested by bond-charge dimerization along the 
$a$-axis together with optical second-harmonic generation and photoinduced CO 
melting.\cite{Yamamoto08,Tanaka10,Nakaya10} Our data can be most naturally 
attributed to the motion of charged kink-type defects -- solitons or domain 
walls in the charge order texture. Both domain walls and solitons stand for 
short wavelength excitations; however whereas a soliton is usually a 
one-dimensional object, the domain wall is not dimensionally restricted. Charge 
neutrality constraint of the CO in \alphaETI{} (a change of stripes equivalent 
to strictly replacing unit cells of one twin type with another) suggests two 
types of solitons and/or domain walls. The first one is the domain wall in pairs 
(a soliton-antisoliton pair) between CR and CP stripes along the $b$-axis, which 
we get if we impose the constraint along the $b$-axis [Fig.\ \ref{fig:domainwalls}(a)].
The second type of domain-wall pair is given by applying the constraint along 
the $a$-axis so that the domain walls' interior contains both charge signs 
[Fig.\ \ref{fig:domainwalls}(b)]. The motion of such entities induces a 
displacement current and can therefore be considered as the microscopic origin 
of polarization in the CO state. Namely, in the presence of an external electric 
field perpendicular to the horizontal stripes, $\mathbf{E}\parallel a$, coupling 
to the AA$^\prime$ dipole moments of each unit cell breaks the symmetry between 
the two orientations of the dipole. Due to first-neighbor interactions the 
AA$^\prime$ dipoles can most easily be flipped at the domain wall, causing the 
wall pairs to move. Coupling to a field in the $\mathbf{E}\parallel b$ 
configuration, parallel to stripe direction and therefore perpendicular to the 
AA$^\prime$ dipoles, seems to be more troublesome. However, one only needs to 
remember that A and A$^\prime$ also interact with B and C sites. At the 
energetically unfavorable domain wall the B and C molecules effectively couple 
AA$^\prime$ dipoles to perpendicular external fields and allow for solitonic 
motion along the $b$-axis as well.

Theoretically, a domain wall should be independent on position, however in a 
real crystal it gets pinned to defect sites.\cite{Chaikinbook} Such a pinning 
causes the domain wall to sit in a local energy minimum. The broad distribution 
of relaxation times might then be ascribed to a distribution of 
activation energies associated with pinning sites. A weak ac electric field 
induces a dielectric response which can be attributed to the activation between 
different metastable states over energies barriers. These metastable states 
correspond to local changes of charge distribution across the length scale of 
domain wall thickness. Finally, a nearly temperature-independent mean relaxation 
time indicates that resistive dissipation cannot be dominant for domain wall 
pairs and that the dielectric relaxation is governed by low energy barriers.

At the end, we consider alternative microscopic descriptions of excitations 
which could give rise to the dielectric relaxation in \alphaETI{}. A model 
proposed by Hotta \etal{}\cite{Hotta08} describes single-particle excitations in 
an insulating, horizontally-striped charge order on a triangular lattice using 
plaquette processes. These correspond to a localized charged particle moving 
between charge-rich and charge-poor stripes. However, the model does not lend 
itself easily to workable predictions of transport, dielectric and infrared 
properties. Another single-particle description  might be found in the excitonic 
model used by Yamaguchi \etal{} to explain dielectric properties below $T=2$~K 
as well as nonlinear conductivity of $\theta$-BEDT-TTF$_2M$Zn(SCN)$_4$ 
($M=\textrm{Cs, Rb})$.\cite{Takahide06,Takahide10} This model considers a 
charge-ordered quarter-filled square-lattice system in which excitations are created 
by moving a localized hole to a site where no hole was originally present, thus 
creating a pair of a localized electron and hole. Due to their electric fields 
being confined to the polarizable plane of ET molecules, the attractive Coulomb 
potential between a bound electron and hole is of a logarithmic form: the 
potential is modeled as $U=U_0 \ln{(r/a)}$ for $r<\lambda$, and a constant 
$U=U_0 \ln{(\lambda/a)}$ for $r>\lambda$, where the spatial scale $a$ is the 
distance between neighboring BEDT-TTF sites, $a \approx 0.5$~nm, and $\lambda$ 
is the screening length. At finite temperatures bound pairs are thermally 
excited. The potential barrier for a pair to unbind under an electric field $E$ 
is given by $2\Delta(E) \approx U_0 \ln{\{U_0/(eEa)\}}$, $E \ll U_0/(ea)$. In 
the excitonic picture a transport gap $2\Delta(E=0) = U_0 \ln{(\lambda/a)}$ 
naturally follows from the cutoff of logarithmic potential.\cite{Takahide10} The 
thermally excited but bound pairs are polarized in the presence of an electric 
field and give rise to a temperature-dependent dielectric constant
\begin{multline}
\varepsilon(T;\omega=0) = 1 + \frac{n_0}{\varepsilon_0} \int_0^\lambda r \ud r \frac{(er)^2}{2k_\mathrm{B}T} \times \\
\left. \times \exp\left\{-\frac{U(r)}{k_\mathrm{B}T}\right\} \middle/ \int_0^\lambda r\ud r \exp \left\{-\frac{U(r)}{k_\mathrm{B}T}\right\} \right.,
\label{eq:exciton-eps}
\end{multline}
where $n_0$ is the electron-hole density at $T \to \infty$ which we take to be 
equal to half the BEDT-TTF density.\cite{Takahide10} The above expression can be 
compared with the measured dielectric properties of \alphaETI{}. Namely, the 
total dielectric constant [$\varepsilon(\omega=0)$] is well-approximated by the 
$\Delta\varepsilon$ of LD mode (see Figs.\ \ref{fig:lfdsparamsdiag} and 
\ref{fig:lfdsparamsab}). A two-parameter ($\lambda$, $U_0$) fit to, \eg{}, our 
$\mathbf{E}\parallel a$ data above $T=25$~K reproduces adequately the general 
temperature dependence, but gives a rather small $\lambda=2.2\pm0.1$~nm and 
$U_0=1.6\pm1.5$~meV. The fit value of $U_0$ is in stark contrast with the value 
extracted from transport gap, $U_0 = \Delta / \ln(\lambda/a) \approx 50$~meV 
(further, substituting $\lambda$ from the previous expression gives an 
unsatisfactory one-parameter fit). A similar discrepancy can be seen in 
$\mathbf{E}\parallel [1 \bar{1} 0]$ and $\mathbf{E}\parallel b$ fits. Also, it 
is not clear whether this model (here shown only in the static limit) reproduces 
the shapes of our experimental dielectric spectra and their temperature 
dependence. In the end the excitonic picture, while nicely applicable to 
$\theta$-BEDT-TTF$_2M$Zn(SCN)$_4$ below 2~K, does not seem to account for 
general dielectric features of \alphaETI{} in the charge-ordered phase. 

\section{Summary}
We investigated electrodynamic properties in the single-crystals of the 
layered organic compound \alphaETI{}. In the normal phase, we observe an 
overdamped Drude response and a weak optical conductivity anisotropy. This is 
consistent with an almost isotropic, weakly temperature-dependent dc 
conductivity inside the conducting layers. Broad intramolecular vibrations of 
the BEDT-TTF molecule might be attributed to charge ordering fluctuations which 
form at short length scales. We demonstrate the abrupt onset of static charge 
order below $T_\mathrm{CO}=136$~K followed by a dramatic drop of the optical 
conductivity. The charge diproportionation remains constant on further cooling. 
The observed charge values are +0.8 and $+0.85e$ on charge-rich sites, and +0.2 
and $+0.15e$ on charge-poor sites, consistent with the charges estimated by 
x-ray. Below the charge-order transition we detect the strong development of 
in-plane dc conductivity and dc gap anisotropy in contrast to the anisotropy of 
optical conductivity which remains weak and similar to high temperatures. The 
optical gap is approximately 75~meV. The development of dc conductivity 
anisotropy in the conducting layers is accompanied by appearance of two 
dielectric relaxation modes in kHz--MHz frequency range. The large dielectric 
mode features an anisotropic phason-like behavior, whereas the small dielectric 
mode is temperature-independent and its properties are reminiscent of a 
soliton-like behavior. All these results make it rather clear that the most 
consistent picture of the horizontal stripe, charge-ordered state in layered 
\alphaETI{} is a cooperative bond-charge density wave with ferroelectric-like 
nature, rather than a fully localized Wigner-crystal. Additional theoretical and 
experimental work is warranted in order to refine the microscopic description of 
both metallic and insulating phase of \alphaETI{}. The emerging broader issue of
the anisotropic, screened phason-like dispersion, its associated
sinusoidal charge-density modulation and whether they are a general signature
of charge order in layered strongly correlated systems certainly deserves
further efforts.

\begin{acknowledgments}
We thank G.\ Untereiner for the sample preparation and T.\ Vuleti\'{c} for his 
help in data analysis. In addition, we wish to acknowledge the contribution of 
B.\ Gorshunov to the experimental part of this study. N.D. is grateful for the 
support by the Magarete-von-Wrangell-Pro\-gramm of Baden-W\"{u}rttemberg. This 
work was supported by the Croatian Ministry of Science, Education and Sports 
under Grants No.\ 035-0000000-2836 and 035-0352843-2844 and by the Deutsche 
Forschungsgemeinschaft (DFG) under Grant DR 228/29-1. We also express our thanks 
to S.\ Mazumdar, T.\ Yamaguchi, C.\ Hotta, and S.\ Brown for enlightening
discussions.
\end{acknowledgments}


\begin{thebibliography}{99}
\bibitem{note2}
The maxima in optical conductivity are obscured by the antiresonance at these
frequencies, leaving only two side peaks remaining.
\bibitem{note3}
The difference to Ref.\ \onlinecite{Dressel94} (where $2\Delta\approx 400$~cm{} is
given) is mainly due to the way how the gap value was determined.
\end{thebibliography}
\end{document}